\documentclass[useAMS,usenatbib]{mnras}
\usepackage{graphicx}
\usepackage{longtable}
\usepackage{amssymb}
\usepackage{amsmath}
\title[X-ray weakness in BALQSOs]{No evidence for an Eddington-ratio dependence of X-ray weakness in BALQSOs }
\author[F. Vito et al.]
{F. Vito,$^{1,2}$\thanks{E-mail: fvito@psu.edu}
W.N. Brandt,$^{1,2,3}$
B. Luo,$^{4,5}$
O. Shemmer,$^{6}$
C. Vignali,$^{7,8}$ and
R. Gilli$^{8}$
\\  \\ 
$^{1}$ Department of Astronomy \& Astrophysics, 525 Davey Lab, The Pennsylvania State University, University Park, PA 16802, USA\\
$^{2}$ Institute for Gravitation and the Cosmos, The Pennsylvania State University, University Park, PA 16802, USA\\
$^{3}$ Department of Physics, The Pennsylvania State University, University Park, PA 16802, USA\\
$^{4}$ School of Astronomy and Space Science, Nanjing University, Nanjing 210093, China\\ 
$^{5}$Key Laboratory of Modern Astronomy and Astrophysics (Nanjing University), Ministry of Education, Nanjing, Jiangsu 210093, China\\
$^{6}$ Department of Physics, University of North Texas, Denton, TX 76203\\
$^{7}$ Dipartimento di Fisica e Astronomia, Universit\`a degli Studi di Bologna, via Gobetti 93/2, 40129 Bologna, Italy \\
$^{8}$ INAF -- Osservatorio di Astrofisica e Scienza dello Spazio di Bologna, via Gobetti 93/3, 40129 Bologna, Italy\\
}

\newcommand*{\lunit}{\ensuremath{\mathrm{erg\,s^{-1}}}}
\newcommand*{\funit}{\ensuremath{\mathrm{erg\, cm^{-2}s^{-1}}}}
\newcommand*{\xmm}{\textit{XMM-Newton}}

\newcommand*{\nhunits}{\mathrm{cm^{-2}}}

\newcommand{\angstrom}{\mbox{\normalfont\AA}}
\newcommand{\daox}{\mbox{$\Delta\alpha_{ox}$}}

\begin{document}
\date{}
\graphicspath{{.}}
\pagerange{\pageref{firstpage}--\pageref{lastpage}} \pubyear{2018}
\maketitle
\label{firstpage}

\begin{abstract}

Several works have studied the relation between X-ray, UV, and wind properties in broad absorption line quasars (BALQSOs), generally concluding that the formation of strong winds is tightly connected with the suppression of the ionizing EUV/X-ray emission. The Eddington ratio ($\lambda_{Edd}$), which measures the accretion rate, is also known to be related with outflow and emission-line properties in the general quasar population. Moreover, models describing quasar accretion depend on $\lambda_{Edd}$, which can thus possibly affect the relative production of accelerating UV and ionizing EUV/X-ray radiation. In this work, for the first time, we investigated whether BALQSO X-ray properties are related with the Eddington ratio. We selected a sample of 30 BALQSOs with accurate measurements of black-hole mass and BAL properties from the literature, and we complemented it with 4 additional BALQSOs we observed with \xmm\, to populate the low and high Eddington-ratio regimes. We did not find evidence for a strong relation between $\lambda_{Edd}$ and X-ray suppression, which however shows a significant correlation with the strength of the UV absorption features. These findings are confirmed also by considering a sample of mini-BALQSOs collected from the literature.

\end{abstract}


\section{Introduction}\label{intro}

One of the most outstanding pieces of evidence for the existence of AGN-driven outflows is the typical broad \mbox{($>2000\,\mathrm{km\,s^{-1}}$)} absorption features visible in the UV spectra of Broad Absorption Line QSOs (BALQSOs), which account for $\sim15\%$ of optically-selected QSOs \citep[e.g.][]{Trump06,Gibson09a}. Such absorption features often have complex structures and are blueshifted with respect to the rest-frame line wavelength, implying outflowing velocities up to $\sim0.2c$ \citep[e.g.][]{Rogerson16}. Less-extreme features ($1000-2000\,\mathrm{km\,s^{-1}}$ in width) are present in optical/UV spectra of the so-called mini-BALQSOs, whose number is comparable to or even greater than the number of BALQSOs, demonstrating the widespread presence of outflows among the whole quasar population \citep[e.g.][]{Ganguly08}.  

The origin of such outflows is thought to be connected to the formation of equatorial winds radiatively driven by UV-line pressure, launched from the accretion disk at \mbox{$\sim10^{16-17}\mathrm{cm}$} \citep[e.g.][]{Proga00}. The ``accretion-disk-wind" model requires the outflowing material not to be over-ionized, as the line-driving efficiency drops when the ionization state 
of the wind is too high. Several hypotheses have been proposed to avoid such over-ionization, spanning from the presence of shielding material (perhaps a failed wind; e.g. \citealt{Proga04}) located at the base of the wind that absorbs the EUV/X-ray radiation emitted from the inner regions of the disk, to a high density of the wind itself due to radiation-pressure confinement \citep[e.g.][]{Baskin14}, to occasional {\it intrinsic} (i.e. not due to absorption) EUV/X-ray weakness \citep[e.g.][]{Luo13,Luo14}, as predicted by some accretion models during phases of fast accretion \citep[e.g.][and references therein]{Meier12,Jiang18}.

Some of these scenarios predict an observed  X-ray weakness (i.e. a weaker observed X-ray emission than the level expected from the UV luminosity), estimated with the \mbox{$\Delta\alpha_{ox}=\alpha_{ox}(\mathrm{observed})-\alpha_{ox}(L_{2500\angstrom})$} parameter, where $\alpha_{ox}(\mathrm{observed})=0.3838\times\mathrm{Log(L_{2keV}/L_{2500\angstrom})}$
 is the slope of a power-law connecting the UV at 2500 $\angstrom$ and the observed \mbox{X-ray} at 2 keV. The value of $\alpha_{ox}$ has been found to correlate with $L_{2500\angstrom}$ (e.g. \citealt{Vignali03,Strateva05, Steffen06}, \mbox{\citealt{Just07,Lusso10}}) up to $z\approx6$ \citep{Nanni17}, and $\alpha_{ox}(L_{2500\angstrom})$ is thus the expected value inferred from the UV luminosity. The quantity $\Delta\alpha_{ox}$ therefore quantifies the deviation of the observed \mbox{X-ray} luminosity with respect to the expectation. 
 
 Indeed, BALQSOs have been generally found to be \mbox{X-ray} weak by up to a factor of  $\sim 100$ ($\Delta\alpha_{ox}\sim-0.75$, e.g. \citealt{Gibson09a}). 
 To discriminate absorption from {\it intrinsic} X-ray weakness as the cause of the {\it observed } X-ray weakness, emission in rest-frame hard 
 X-rays, which are not affected by low-to-moderate column densities of absorbing material, must be studied. For instance, \cite{Luo14} using \textit{NuSTAR} data found a significant fraction of intrinsically X-ray weak BALQSOs in their local sample. Other authors \citep[e.g.][]{Gallagher06,Morabito14,Liu18} accessed high rest-frame energies using samples of high-redshift ($1.4<z<2.9$ ) BALQSOs observed with either \textit{Chandra} or \xmm. In particular, \cite{Morabito14} estimated an average intrinsic X-ray weakness of a factor of $\approx3$ ($\Delta\alpha_{ox}\approx-0.2$). Recently, \cite{Liu18} found a fraction of intrinsically X-ray weak BALQSOs of $\approx6-23\%$ among their $z=1.6-2.7$ sample, significantly higher than the $\lesssim2\%$ fraction of X-ray weak quasars among the general non-BALQSO population \citep[e.g.][]{Gibson08}. However, results based on X-ray spectral analysis in many cases reveal the presence of X-ray absorption \citep[e.g.][]{Gallagher02, Grupe03, Shemmer05, Giustini08}.
 A number of correlations are known between the X-ray weakness and other observational properties of BALQSOs, such as the minimum and maximum velocity of the outflow, and the strength of the absorption features \citep[e.g.][]{Gallagher06, Fan09,Gibson09a,Wu10}, suggesting that the level of \mbox{X-ray} emission indeed has material effects in shaping the wind observed in the UV.

  BALQSOs are generally thought to be powered by fast-accreting SMBHs \citep[e.g.][]{Boroson02, Meier12}, where the accretion rate is measured through the Eddington ratio, defined as $\lambda_{Edd}=L_{bol}/L_{Edd}$, where \mbox{$L_{Edd}=1.26\times10^{38}M_{BH}/M_\odot\,\mathrm{erg\,s^{-1}}$}. In fact, while BALQSOs have been found to have $\lambda_{Edd}$ as low as $\approx0.1$, the fraction of quasars showing BAL features increases with $\lambda_{Edd}$ \citep{Ganguly07}. However, as the accretion rate increases, a larger amount of ionizing EUV radiation is produced according to standard accretion models \citep[e.g.][]{Shakura73}. One possibility to avoid overionization of the outflowing material in fast-accreting BALQSOs (i.e. $\lambda_{Edd}\approx1$) is that suppression (either intrinsic or due to absorption) of the EUV-to-X-ray emission, responsible for ionization, depends on the Eddington ratio. In this case, one may expect an anti-correlation between \daox\, and $\lambda_{Edd}$, i.e. a change of the typical observed (and perhaps intrinsic) spectral shape of BALQSOs approaching the Eddington limit. Hints for such a scenario have been derived by, e.g. \cite{Lusso10,Lusso12} by studying the dependence of $\alpha_{ox}$ and the quasar bolometric correction on $\lambda_{Edd}$ (but see also, e.g. \citealt{Plotkin16}). From a theoretical point of view, the \cite{Shakura73} accretion mode cannot be in place for Eddington ratios exceeding or even approaching unity. Several models have been proposed to describe Eddington-limited accretion flows, some of which predict an intrinsic suppression of X-ray photon production \citep[e.g.][and references therein; but see also, e.g. \citealt{CastelloMor17}]{Meier12,Jiang18}. We additionally note that the UV part of quasar SEDs, which is responsible for line-pressure acceleration, may also depend on the accretion rate. Increased UV line pressure may compete against EUV/X-ray ionization in the shaping of quasar wind properties \citep[e.g.][]{Kruczek11, Richards11}, although no significant evidence for a strong variation of the optical/UV part of quasar SEDs for different regimes of $\lambda_{Edd}$ has been derived observationally \citep[e.g.][]{Scott14}.
  
  The Eddington ratio is considered one of the fundamental parameters driving observable quasar properties \citep[e.g.][]{Shen14}, and it is indeed closely related with the quasar ``Eigenvector 1" \citep[e.g.][]{Boroson02b}, i.e. a preferred direction in the quasar multidimensional parameter space along which quasar emission-line properties are aligned. BALQSOs appear to follow the same relations with $\lambda_{Edd}$ as the overall quasar population \citep[e.g.][]{Yuan03}. Moreover, outflow properties (e.g. velocity) in quasars \citep[e.g.][]{Marziani12} and, in particular, BALQSOs \citep[e.g.][]{Ganguly07} have been found to correlate with $\lambda_{Edd}$. However, a possible dependence of the X-ray properties of BALQSOs on accretion rate has never been investigated. 
In this work, we made use of both archival and proposed X-ray observations of a sample of 34 high-redshift ($1.5\lesssim z \lesssim 2.2$) SDSS BALQSOs with accurate measurements of black-hole mass to study the dependence of the \textit{observed} X-ray weakness on Eddington ratio. We used an $H_0=70\,\rmn{km\,s^{-1}Mpc^{-1}}$, $\Omega_m=0.3$, and $\Omega_\Lambda=0.7$ $\Lambda$CDM cosmology.

\section{Sample selection and analysis}

The goal of this work is to study the observed X-ray emission level relative to the UV luminosity, in terms of \daox, as a function of $\lambda_{Edd}$ for a sample of high-redshift BALQSOs, where the redshift range ($1.5\lesssim z \lesssim 2.2$) was chosen such that both the Mg II and C IV emission lines are included in the SDSS spectral coverage, the former to derive SMBH masses, and the latter to detect the BAL features. By imposing quality cuts on the signal-to-noise ratio ($SNR$) of the observed SDSS spectra, we limited our study to those objects with securely identified BAL features (i.e. which cannot be ascribed to noisy spectra) and accurate measurements of black-hole mass and BAL properties (e.g. absorption-line strength).
In this section, we describe the sample of BALQSOs observed in the X-rays we collected from the literature, and the sample of four BALQSOs we observed with \xmm.
%For the latter sources, we also describe the data reduction and analysis. 

\subsection{Selection of the BALQSO sample from the literature}\label{lit_sample}

\cite{Fan09}, \cite{Gibson09a}, and \cite{Morabito14} presented the X-ray properties of three samples of 41, 73, and 18 BALQSOs, respectively, for a total of 108 unique objects, covered by \textit{Chandra} or \textit{XMM-Newton} observations. We matched them with the catalog of virial black-hole masses and bolometric luminosities for SDSS DR7 QSOs of \cite{Shen11}. %In order to limit our analysis to sources with accurate measurements of BAL properties, w
In order to study a homogeneous sample of objects, representative of the majority of the BALQSO population, and to avoid additional complexity due to different ionization properties of the outflowing material, we considered only objects flagged as high-ionization BALQSOs (HiBALQSOs), thus discarding 9 objects classified as low-ionization BALQSOs (7) or even non-BALQSOs (2) in \cite[who used more recent SDSS spectra than \citealt{Gibson09a} and \citealt{Fan09}]{Shen11}. 
	
The X-ray emission produced at the base of the jets in radio-detected QSOs is known to be comparable to or even dominant over the disk/corona-linked X-ray emission \citep[e.g.][]{Miller11}, which is the physical mechanism of interest in this study. An additional X-ray contribution
from the jets would thus artificially increase the observed X-ray flux/luminosity leading to biased estimates of $\alpha_{ox}$ and $\Delta\alpha_{ox}$. Disentangling the two contributions (corona vs. jets) requires a careful spectral analysis, which is prevented by the small number (up to few tens, with a median number of counts of 7) of X-ray counts for the sources considered in this work. Excluding quasars detected in large-area radio surveys from our sample avoids this bias. 	
We therefore discarded 15 radio sources detected in the FIRST radio catalog \citep{White97}, and 2 QSOs not covered by the FIRST survey, as flagged in \cite{Shen11}, resulting in a parent sample of 82 sources. 

Spectral noise can sometimes mimic the absorption features affecting the C IV emission line used to define BALQSOs. In order to select a clean sample of BALQSOs, we imposed a quality cut on the SDSS spectra, requiring a $SNR$ at wavelengths close to the C IV emission line of $SNR_{CIV}>5$ (10 objects discarded). This requirement also ensures an accurate measurement of the balnicity index (see \S~\ref{results}), an indicator of the absorption strength, which we use later in the analysis together with the Eddington ratio to investigate the dependencies of the X-ray weakness. 

The reliability of the Eddington-ratio estimates is strongly dependent on the accuracy of the measured black-hole mass. Single-epoch virial black-hole masses are usually estimated through scaling relations with the FWHM of the Balmer (H$\alpha$ and H$\beta$), Mg~II, and C~IV emission lines, in order of reliability (see the detailed discussion in \citealt{Shen13}; see also, e.g. \citealt{Shen11,Kozlowski17}). The use of the Balmer emission-line series is precluded to us by the need for spectral coverage of the C IV line, necessary to detect the BAL features. The same absorption lines can affect strongly the shape of the C~IV emission lines, and thus the measurement of the FWHM,  preventing the use of the C~IV emission line to estimate black-hole masses for our sample. We therefore use virial black-hole masses derived from the Mg~II emission line consistently for all of our sample. The required simultaneous spectral coverage of the Mg~II and C~IV lines restricts the redshift range of this work to $1.5\lesssim z \lesssim2.2 $, thus discarding an additional 18 BALQSOs for which Mg~II-based black-hole masses are not available.

We finally impose quality cuts on the Mg~II line detection ($SNR_{MgII}>5$) and fit  ($\chi^2_{MgII}<1.2$) to include only sources with accurate measurements of black-hole mass, further restricting the sample to 30 BALQSOs. We report in Tab.~\ref{tab_selection} a summary of the number of sources surviving each of the selection steps, and in  Tab.~\ref{tab_sample} the properties of the final 30 selected sources.
Our conclusions below hold if more conservative quality cuts are applied (e.g. $SNR_{CIV}>10$ and $SNR_{MgII}>10$), at the cost of greatly reducing the sample size, as discussed in \S~\ref{results}.

\begin{table}
	\caption{Summary of the number of X-ray sources collected from the literature surviving after each selection step (as described in \S~\ref{lit_sample}), starting from the parent sample of 108 unique objects included in \citet{Fan09}, \citet{Gibson09a}, and \citet{Morabito14}.}
	\begin{tabular}{|cc|} 
		\hline
		Parent sample & 108 \\
		HiBAL & 99 \\
		Radio undetected & 82 \\
		$SNR_{CIV}>5$ & 72 \\
Spectral coverage of Mg~II 	& 54 \\
	$SNR_{MgII}>5$ and $\chi^2_{MgII}<1.2$ & 30\\
	\hline
	\end{tabular}  \label{tab_selection}\\
\end{table}

We computed the monochromatic UV luminosities from  the flux at (rest-frame) $2500\,\angstrom$  derived by \cite{Shen11} through spectral fitting, and homogeneously applied the \cite{Steffen06} calibration, considering $L_{2500\angstrom}$ as the independent variable:
\begin{equation}
\alpha_{ox}=-0.137\times\mathrm{log}L_{{2500\angstrom} }+ 2.638
\end{equation}
\cite{Steffen06} derived $\alpha_{ox}(L_{2500\angstrom})$ for a sample of optically-selected quasars. Since ours is an SDSS-selected sample, we preferred \cite{Steffen06} over \cite{Lusso10}, who applied an X-ray selection. However, the two calibrations return almost exactly the same results for our sample. We also retrieved the observed luminosities at 2 keV ($L_{2\,\mathrm{keV}}$) provided by the original works to compute $\alpha_{ox}$(observed).

Most of the selected BALQSOs are included in the \cite{Gibson09a} sample, and a few of them are in common with either \cite{Fan09} or \cite{Morabito14}. Two BALQSO are selected from the \cite{Fan09} compilation only, while none is included in the \cite{Morabito14} sample only (see last column of Tab.~\ref{tab_sample}). When \mbox{X-ray} luminosities for a quasar are provided by more than one author, we consistently assumed the value reported by \cite{Gibson09a}, from which the majority of the sample is selected. X-ray luminosities are not corrected for absorption. The general paucity of counts ($\lesssim80$, with a median value of 7) prevents a detailed spectral analysis, which would be required to estimate with acceptable accuracy the intrinsic luminosities of these sources. In fact, complex absorption models (e.g. with ionized or partially covering absorption) may be suitable descriptions of BALQSO X-ray spectra \citep[e.g.][]{Gallagher02}, and cannot be well constrained with the available small number of counts.

Fig.~\ref{fig1} presents the observed X-ray weakness as a function of Eddington ratio for these 30 sources as circles.
A Spearman's $\rho$ test and a generalized Kendall's $\tau$ test\footnote{We used the ASURV v1.2 package (http://www.astrostatistics.psu.edu/statcodes/sc\_censor.html), which accounts for censored data.} returns a probability in favor of a correlation between \daox\, and $\lambda_{Edd}$ of 0.9977 and 0.9983, respectively. These values are a hint for a possible correlation, but the sample does not populate the extreme tails of the Eddington-ratio distribution; i.e. $\lambda_{Edd}\lesssim0.1$ and, especially, $\lambda_{Edd}\approx1$, preventing an accurate assessment of a possible relation.

True uncertainties on the estimated black-hole masses are due to a complex combination of factors, like the measurement of the line profile and width, and the continuum luminosity. \cite{Shen11} report errors on black-hole masses which include these observational measurement errors. However, the dominant uncertainty factor is the scatter of the calibration between black-hole mass and Mg~II line width, which is $\approx0.35$ dex \citep[e.g.][]{Shen11}. In Fig.~\ref{fig1} we use this value to estimate the uncertainty on $\lambda_{Edd}$. We also show the median error on $\Delta\alpha_{ox}$ derived from the statistical uncertainties on the X-ray counts only. We note that the non-parametric statistical tests we used are based on point ranking, and thus are not nominally sensitive to the error bars, although their statistical power decreases in the case of large uncertainties. 

\begin{table*}
	\caption{Main properties of the sample of BALQSOs selected from the literature and BALQSOs newly observed in X-rays.}
	\begin{tabular}{|cccccccc|} 
		\hline
		\multicolumn{1}{|c|}{{ SDSS ID }} &
		\multicolumn{1}{|c|}{{ $z$}} &
		\multicolumn{1}{|c|}{{ log$(L_{bol})$}} &
		\multicolumn{1}{|c|}{{ log$(M_{BH}/M_\odot)$}} &
		\multicolumn{1}{|c|}{{ log$\lambda_{Edd}$}} &
		\multicolumn{1}{|c|}{ \daox} & 
		   \multicolumn{1}{|c|}{ $\frac{\mathrm{BI}_0}{\mathrm{km\,s^{-1}}}$} &
		\multicolumn{1}{|c|}{ Ref.} \\ 
		\hline
	 \multicolumn{8}{|c|}{Sources studied in previous works} \\	
		011227.60$-$011221.7    &1.76  &47.01   &9.51    &  $-0.60 $ & $-0.70$ & 2150 &G09\\		
		020230.66$-$075341.2  &1.72  &46.60 & 9.12   &  $-0.62$  & $<-0.24$ & 208&F09\\
		024304.68+000005.4 &1.99  &46.95 & 9.70    & $-0.85$  & $-0.10$& 325 & F09, G09\\
		083104.90+532500.1  &2.07  &47.40 & 10.00  & $-0.70$   &$-0.51$ &556&  G09\\
		084538.66+342043.6 &2.15  &47.59 & 10.27   & $-0.79 $  &$-0.18$ & 2638 & G09 \\				
		092138.44+301546.9  &1.59  &46.85  &9.34   & $-0.59$   &$-0.51$ & 1 & F09, G09\\
		093514.71+033545.7 &1.82   &47.54    & 9.35    & $+0.09 $    & $-0.38$ &5 & G09, M14\\			
		094309.55+481140.5 & 1.81  & 46.77  & 9.17    & $-0.50$ & $<-0.33$ &  52.2 & G09,F09 \\
		094440.42+041055.6  &1.98 &46.71   &8.88   & $-0.27$  & $<-0.34$  & 1104 & F09\\
		095944.47+051158.3  &1.60  &46.52 & 9.71     & $-1.29$    &$-0.15$ & 150 & G09\\
		100711.80+053208.9  &2.10   &47.81    &10.43    & $-0.72$    & $-0.27$ & 4263 & G09, M14\\	
		110637.15+522233.3  & 1.84  & 46.58  & 9.33   & $-0.85$   & $-0.26$ &  588.0 & G09 \\	
		120522.18+443140.4  & 1.92 & 46.63  & 9.48   & $-0.95$  &  $-0.23$   &  1059.1 & F09, G09\\		
		120626.17+151335.5   &1.63  &46.79  & 9.30    & $-0.61 $   &$-0.47$ & 1429 & G09\\
		121125.48+151851.5    &1.96  & 46.75 & 9.45   & $-0.80 $    &$-0.41$ & 5137 & G09\\
		121440.27+142859.1   &1.62  &47.14   &9.37   & $-0.33$      &$<-0.62$ & 3301 & G09\\
		121930.95+104700.0  &1.62  &47.09  & 9.44    & $-0.44$     &$-0.49$ & 3288 & G09\\
		122637.02+013016.0  & 1.55 & 46.86 & 9.58  & $-0.82$      &  $-0.27$ & 3159.7 & G09\\	
		130136.12+000157.9   &1.78  &47.16   & 9.83    & $-0.77$   &  $-0.42$ & 4522 & G09 \\
		142620.30+351712.1  & 1.75 & 46.7   & 9.31      & $-0.71$    &  $0.02$ & 245.2 & G09\\	
		142640.83+332158.7  & 1.54  & 46.34 & 9.36  & $-1.12$   &  $-0.31$ & 101.5 & G09\\
		142652.94+375359.9  & 1.81  & 46.54 &9.56  &$-1.13$ & $-0.26$      &  239.5& G09,F09 \\
		143031.78+322145.9 & 2.21  & 46.46  & 9.6  & $-1.24$ & $<0.16$   &  308.8 & G09\\	
		143117.93+364705.9   & 2.1  & 46.84  & 9.42  & $-0.68$ & $-0.05$  &  426.1 & G09\\	
		143411.23+334015.3  & 1.79  & 46.48  & 9.5  & $-1.12$  &  $-0.03$ & 70.7 & G09 \\
		143513.90+484149.2 & 1.89 & 46.74  & 9.62  & $-0.98$ &  $0.01$ &  314.2 & G09\\
		143752.75+042854.5 & 1.92 & 47.05 & 9.97   &  $-1.02$    & $<-0.42$ & 2663 & G09, M14\\
		143853.36+354918.7 & 1.55 & 46.51  & 9.04  & $-0.63$     &  $-0.14$  &  1079.8 & G09 \\
		155338.20+551401.9  &1.64  &46.72   & 9.69     &$-1.07$    & $-0.22$ & 18 &G09 \\
		235253.51$-$002850.4 &1.62  &46.83   & 9.35    & $-0.62$     &$<-0.82$ & 4307 &G09\\
\hline
	 \multicolumn{8}{|c|}{Newly observed sources} \\
	 0938+3805 & 1.828 & 47.51 & 9.4 & 0.01 &$-$0.08 & 167 & This work\\ 
	 1112+0053  &1.687  &47.07 &  10.2 &  --1.22 & $-$0.09 &440 & This work\\ 
	 1252+05273 &  1.900 & 47.42&   9.5&   --0.21 & 0.12 &95 & This work\\ 
	 1644+4307&1.715  &47.28 &  10.1 & --0.96 & $-$0.14 & 85 & This work\\ 	
		\hline
	\end{tabular}  \label{tab_sample}\\
IDs, redshifts, bolometric luminosities, and SMBH masses are from \citet{Shen11}. Bolometric luminosities are computed from the rest-frame $3000\angstrom$ or $1350\angstrom$ luminosities at $z<1.9$ and $z>1.9$, respectively.  Virial SMBH mass estimates are derived from the Mg II emission line. See  \citet{Shen11} for details. {We computed the Eddington ratios using the bolometric luminosities and black-hole masses reported in this table.} Values of \daox\, are computed as described in \S~\ref{lit_sample} for sources with previous X-ray observations and in \S~\ref{xmm_obs} for newly observed sources. Values of the balnicity indices ($\mathrm{BI}_0$) are collected from \citet{Gibson09a}. Last column reports the reference paper for the X-ray observations (F09: \citealt{Fan09}; G09: \citealt{Gibson09a}; M14: \citealt{Morabito14}).
\end{table*}

\subsection{Selection of our \xmm\, targets, data reduction and analysis}\label{xmm_obs}
 In order to increase the number of BALQSOs with low and high values of Eddington ratio, thus allowing us to determine with high significance if the \daox-$\lambda_{Edd}$ correlation exists, during AO16 we obtained \xmm\, observations of four BALQSOs (see Tab.~\ref{new_sample}). They were selected among the BALQSOs included in the \cite{Gibson09a} catalog with black-hole mass from \cite{Shen11} satisfying the quality cuts described in \S~\ref{lit_sample}, with the additional requirement to have log$\lambda_{Edd}\approx1$ or log$\lambda_{Edd}\approx0.1$ to sample better these accretion regimes (see Tab.~\ref{tab_sample}). 
X-ray data were reduced and analyzed following standard SAS\footnote{https://www.cosmos.esa.int/web/xmm-newton/sas-threads} procedures. Periods of high background levels (count rates $>0.4\,\mathrm{cts\,s^{-1}}$ for the PN camera and $>0.35\,\mathrm{cts\,s^{-1}}$ for the MOS cameras) were filtered out. Tab.~\ref{new_sample} reports the filtered exposure times. Our sources have $\approx160-2000$ net counts in the $0.5-10$ keV band, considering the three \xmm\, cameras.

\begin{figure}
	\begin{center}
		\hbox{
			\includegraphics[width=90mm,keepaspectratio]{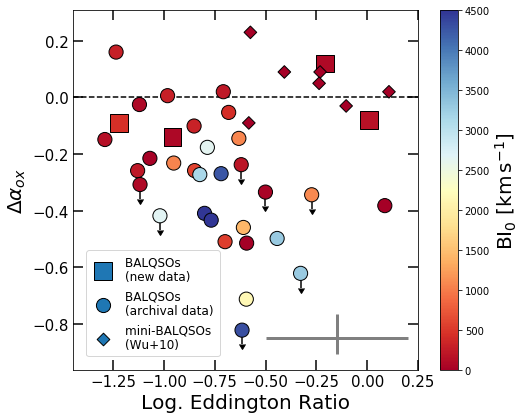} 
		}\vspace{-1cm}
	\end{center}
	\caption{Observed X-ray weakness ($\Delta\alpha_{ox}$) as a function of Eddington ratio for the sample of BALQSOs collected from the literature (circles), our four targets of X-ray observations (squares), and the sample of mini-BALQSOs of \citet[small diamonds]{Wu10}. Downward-pointing arrows represent upper limits ($1\sigma$ confidence level). Symbols are color-coded according to their balnicity index ($BI_0$; see \S~\ref{results}). The horizontal dashed line marks the locus of quasars with normal levels of X-ray emission (i.e. $\daox=0$). The median uncertainties are shown as grey error bars in the bottom-right corner of the plot, and account only for the dominant error factors; i.e. the scatter in the $M_{BH}-FWHM_\mathrm{Mg~II}$ ($\approx 0.35$ dex; e.g. \citealt{Shen11}) relation for $\lambda_{Edd}$, and the statistical uncertainties on the X-ray counts for $\Delta\alpha_{ox}$.}
	\label{fig1}
\end{figure}

To be consistent with the derivation of the observed \mbox{X-ray} weakness in \cite{Gibson09a}, and also to avoid making strong assumptions about the \mbox{X-ray} spectra (e.g. neutral absorption), we performed a spectral analysis assuming a broken power-law model (\textit{wabs$\times$bknpowerlaw} in XSPEC), with break energy fixed at rest-frame 2 keV. 
Best-fitting parameters are reported in Tab.~\ref{spectral_par_bknpow}. For J164452.70+430752.20 we could not constrain the low-energy photon index, which hit the hard low boundary of the allowed parameter range (i.e. $\Gamma=-3$). We therefore fixed it to that value, which hints at a significant level of absorption. 
 X-ray monochromatic luminosities are derived as observed (i.e. no correction for absorption has been applied), similarly to \cite{Gibson09a}. Hereafter, we will use these observed values (squares in Fig.~\ref{fig1}) for consistency with the sample collected from the literature. The utilized model describes the effective (i.e. observed) shape of the X-ray spectrum, and the resulting observed luminosities are thus not strongly dependent on the true intrinsic spectral parameters (e.g. photon index and column density, which for low-to-moderate numbers of counts are strongly degenerate), nor the physical state of the possible absorbing material (e.g. neutral vs. partly ionized).

\subsubsection{Estimating the intrinsic luminosity}
The number of net counts we detected for our targets (see last two columns of Tab.~\ref{new_sample}) allowed us to perform a simple spectral analysis assuming a power-law model with both Galactic and intrinsic neutral absorption (model \textit{wabs$\times$zwabs$\times$powerlaw} in XSPEC\footnote{heasarc.gsfc.nasa.gov/xanadu/xspec/} v12.9.0n, \citealt{Arnaud96}) to estimate their intrinsic photon indices $\Gamma$, column densities, and luminosities. Best-fitting parameters are reported in Tab.~\ref{spectral_par}.  The best-fitting photon indices for two sources are flatter than the common values found for quasars, probably due to their photon counting statistics ($\approx160-230$ net counts, see Tab.~\ref{new_sample}) not being sufficiently high to break the degeneracy between (likely complex) absorption and flat photon index. For these objects we repeated the fit fixing the photon indices to $\Gamma=1.9$. Tab.~\ref{spectral_par} reports also the intrinsic (i.e. absorption-corrected) X-ray monochromatic luminosities, used to compute the \daox\, values, which are thus estimates of the intrinsic X-ray weakness. We note that J164452.70+430752.20 shows the largest intrinsic column density in the sample, as expected from its very hard effective spectrum (see previous section).

We stress that here we make a strong assumption considering the obscuring material neutral and fully covering, and thus for the rest of the analysis, consistently with the sample selected from the literature, we use the observed luminosities and $\Delta\alpha_{ox}$ values derived from the broken power-law model in the previous section. As expected, since here we apply a correction for absorption, the luminosities and  $\Delta\alpha_{ox}$ values reported in Tab.~\ref{spectral_par} are slightly higher than the observed values (Tab.~\ref{spectral_par_bknpow}), although the difference is smaller than the typical $\Delta\alpha_{ox}$ uncertainties. Therefore, the use of observed or intrinsic luminosities for our four sources does not affect significantly the results.

\begin{table*}
	\caption{Summary of \xmm\, observations of our BALQSO targets. Net counts are computed following Appendix A3 of \citet{Weisskopf07}. Exposure times are derived after filtering for background flares.}
	\begin{tabular}{|cccccc|} 
		\hline
		\multicolumn{1}{|c|}{{ SDSS ID }} &
\multicolumn{1}{|c|}{{ OBSID}} &
\multicolumn{1}{|c|}{{ Date}} &
\multicolumn{1}{|c|}{{ $T_{EXP}$}} &
\multicolumn{1}{|c|}{{ $0.5-2$ keV cts}} &
\multicolumn{1}{|c|}{ $2-10$ keV cts} \\ 
		
&\xmm&&ks&PN/MOS1/MOS2& PN/MOS1/MOS2\\
		\hline
093846.80+380549.8 & 0801790401 & 2017-05-01 & 10.6/15.7/14.5 &
 $172^{+14}_{-13}$/$90^{+10}_{-10}$/$124^{+12}_{-11}$ & $57^{+9}_{-8}$/$22^{+5}_{-5}$/$34^{+7}_{-6}$\\
111249.70+005310.10  & 0801790101 & 2017-05-28 & 5.0/11.5/11.5  & $55^{+8}_{-7}$/$39^{+7}_{-6}$/$36^{+7}_{-6}$ & $47^{+8}_{-7}$/$18^{+5}_{-4}$/$31^{+6}_{-6}$\\
125216.60+052737.70 & 0801790601 & 2017-06-18 & 18.2/26.8/26.8 & $656^{+26}_{-26}$/ $246^{+16}_{-16}$/ $270^{+17}_{-16}$  & $160^{+14}_{-13}$/$390^{+21}_{-20}$/$438^{+22}_{-21}$   \\
164452.70+430752.20& 0801790301& 2017-07-13 &  5.0/7.5/7.5   &  $48^{+8}_{-7}$/$18^{+5}_{-4}$/$31^{+6}_{-5}$ &$35^{+7}_{-6}$/$14^{+4}_{-4}$/$20^{+5}_{-5}$\\
		\hline
	\end{tabular}  \label{new_sample}
\end{table*}

%\begin{table}
%	\caption{Main properties of our BALQSO sample}
%	\begin{tabular}{|cccccccc|}
%		\hline
%		\multicolumn{1}{|c|}{{ SDSS ID }} &
%		\multicolumn{1}{|c|}{{ $z$}} &
%		\multicolumn{1}{|c|}{{ log$(L_{bol})$}} &
%		\multicolumn{1}{|c|}{{ log$(\frac{M_{BH}}{M_\odot})$}} &
%		\multicolumn{1}{|c|}{{ log$\lambda_{Edd}$}} & 
%				   \multicolumn{1}{|c|}{ $\frac{\mathrm{BI}_0}{\mathrm{km\,s^{-1}}}$} \\ 
%		\hline
%0938+3805 & 1.828 & 47.51 & 9.4 & 0.01 & 167\\ 
%1112+0053  &1.687  &47.07 &  10.2 &  --1.22 & 440\\ 
%1252+05273 &  1.900 & 47.42&   9.5&   --0.21 & 95\\ 
%1644+4307&1.715  &47.28 &  10.1 & --0.96 & 85\\ 
%		\hline
%	\end{tabular}  \label{new_sample_properties}
%\end{table}

\begin{table*}
	\caption{Best-fitting spectral parameters assuming a broken power-law emission. The luminosity at rest-frame 2 keV is observed (i.e. not corrected for absorption).}
	\begin{tabular}{|ccccccc|} 
		\hline
		\multicolumn{1}{|c|}{{ SDSS ID }} &
		\multicolumn{1}{|c|}{{ $\Gamma_1$}} &
		\multicolumn{1}{|c|}{{ $\Gamma_2$}} &
		\multicolumn{1}{|c|}{{ $F_{0.5-2\,\mathrm{keV}}$}} &
		\multicolumn{1}{|c|}{ $L_{\nu,2\,\mathrm{keV}}$} &
		\multicolumn{1}{|c|}{ \daox}\\ 
		
		&&&$10^{-14}\,\funit$&$10^{27}\,\mathrm{erg\,s^{-1}Hz^{-1}}$ &\\
		\hline
		093846.80+380549.8 &$-0.91^{+0.70}_{-0.81}$ & $1.63^{+0.08}_{-0.08}$ &$3.50$& 1.22 & $-$0.08 \\
		111249.70+005310.10  &$0.67^{+0.22}_{-0.24}$ &$1.80^{+0.27}_{-0.25}$ & 2.36    &  0.61& $-$0.09 \\
		125216.60+052737.70 & $2.06^{+0.17}_{-0.18}$  &  $2.11^{+0.05}_{-0.05}$ &  7.80& 3.48 &  0.12 \\
		%		164452.70+430752.20&$0.24^{+0.26}_{-0.30}$ & $2.03^{+0.43}_{-0.37}$& 2.33&0.34& -0.24\\
		164452.70+430752.20&$-3.00$ fixed & $1.25^{+0.15}_{-0.09}$& 2.29&0.62& $-$0.14\\
		\hline
	\end{tabular}  \label{spectral_par_bknpow}
\end{table*}

\begin{table*}
	\caption{Best-fitting spectral parameters assuming power-law emission absorbed by neutral material. Luminosities are intrinsic.}
	\begin{tabular}{|ccccccc|} 
		\hline
		\multicolumn{1}{|c|}{{ SDSS ID }} &
		\multicolumn{1}{|c|}{{ $\Gamma$}} &
		\multicolumn{1}{|c|}{{ $N_H$}} &
		\multicolumn{1}{|c|}{{ $F_{0.5-2\,\mathrm{keV}}$}} &
		\multicolumn{1}{|c|}{{ $L_{2-10\,\mathrm{keV}}$}} &
		\multicolumn{1}{|c|}{ $L_{\nu,2\,\mathrm{keV}}$} &
		\multicolumn{1}{|c|}{ \daox}\\ 
		
		&&$10^{22}\,\nhunits$&$10^{-14}\,\funit$&$10^{45}\,\lunit$ &$10^{27}\,\mathrm{erg\,s^{-1}Hz^{-1}}$\\
		\hline
		093846.80+380549.8 &$1.90^{+0.13}_{-0.13}$ & $1.4^{+0.4}_{-0.3}$ &$3.59$& $1.51$& 1.81 & $-$0.01 \\
		111249.70+005310.10  & $1.39^{+0.16}_{-0.15}$ & $0.7^{+0.6}_{-0.4}$ &   2.35  & 0.96&  0.73&  $-$0.06 \\
		                                   & 1.90 fixed                     & $2.2^{+0.7}_{-0.5}$ &   2.60  & 1.14 &  1.35 & 0.04  \\
		125216.60+052737.70 &$2.12^{+0.06}_{-0.05}$ &  $<0.1$  &                     7.81   &2.48 &  3.50 & 0.12 \\
		164452.70+430752.20& $1.63^{+0.23}_{-0.21}$ & $2.4^{+1.1}_{-0.9}$&   2.32   & 1.11  &   1.04 & $-$0.05\\
		                                   & 1.90 fixed                     & $3.4^{+0.8}_{-0.7}$ &   2.42  & 1.23 &  1.47 & 0.00  \\		
		\hline
	\end{tabular}  \label{spectral_par}
\end{table*}

\section{Results and discussion}\label{results}
In \S~\ref{daox_lEdd} we describe the results of our investigation of a putative anti-correlation between $\Delta\alpha_{ox}$ and $\lambda_{Edd}$. In \S~\ref{daox_BI0} we study the relation between Eddington ratio and BAL strength, as parametrized by the balnicity index. In \S~\ref{miniBALQSOs} we include in our investigation a sample of mini-BALQSOs that satisfy our selection requirements, in order to expand the analysis to objects with weaker absorption features. Finally, in  \S~\ref{Discussion} we discuss and interpret the results.

\subsection{$\Delta\alpha_{ox}$ versus $\lambda_{Edd}$}\label{daox_lEdd}
Adding our targets with low and high Eddington ratios to the sample retrieved from the literature does not confirm the putative \daox-$\lambda_{Edd}$ anti-correlation. 
In fact, the probability of a correlation decreases to 0.966 and 0.982 according to Spearman's  $\rho$ and Kendall's  $\tau$ tests, respectively.
This is largely due to the two sources with $\lambda_{Edd}\approx1$ showing a level of X-ray emission close to expectation (i.e. \daox$=0$) and much higher than other BALQSOs with similar or slightly lower $\lambda_{Edd}$. We therefore conclude that there is no clear and simple dependence of \mbox{X-ray} weakness on Eddington ratio in BALQSOs.

\subsection{$\Delta\alpha_{ox}$ versus BAL strength}\label{daox_BI0}

Several authors have investigated correlations between \daox\, and the  physical parameters of the outflow in BALQSOs, including the outflow velocity and absorption equivalent width \citep[e.g.][]{Gallagher06,Gibson09a,Wu10}, finding that \daox\, is more negative in BALQSOs with stronger absorption features. We therefore investigated whether our targets follow this trend, and if the Eddington ratio plays a secondary role in driving the X-ray weakness. As a measure of BAL strength, we used the extended balnicity index ($BI_0$), defined by \cite{Gibson09a} as 

\begin{equation}
	BI_0=\int_0^{25000}(1-\frac{f_v}{0.9})Cdv,
\end{equation}

\noindent where $f_v$ is the ratio of the observed spectrum to the continuum model as a function of the velocity $v$, and $C$ is a constant set to unity if the spectrum is at least 10\% below the continuum model for velocity widths of at least $2000\,\mathrm{km\,s^{-1}}$ and zero otherwise. 

\begin{figure}
	\begin{center}
		\hbox{
			\includegraphics[width=90mm,keepaspectratio]{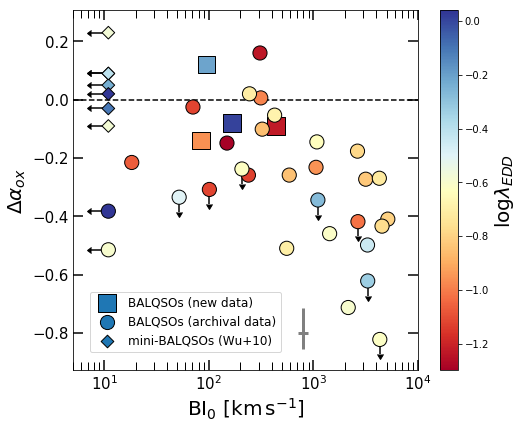} 	
		}\vspace{-1cm}
	\end{center}
	\caption{Observed X-ray weakness ($\Delta\alpha_{ox}$) as a function of $BI_0$. Symbols are the same as in Fig.~\ref{fig1} and are color-coded according to their Eddington ratio. The $x$-axis is in logarithmic units for visual purposes. Mini-BALQSOs (which  have $BI_0=0\,\mathrm{km\,s^{-1}}$ by definition) and BALQSOs with $BI_0<10\,\mathrm{km\,s^{-1}}$ are plotted as upper limits on $BI_0$ (symbols with leftward-pointing arrows). The horizontal dashed line marks the locus of quasars with normal levels of X-ray emission (i.e. $\daox=0$). The median uncertainties are shown as grey error bars in the bottom-right corner of the plot. We used a $10\%$ uncertainty for $BI_0$ \citep[e.g.][]{FilizAk13}, and the statistical uncertainties on the X-ray counts for $\Delta\alpha_{ox}$, which is its dominant factor of uncertainty.}
	\label{fig2}
\end{figure}

Following \cite{Gibson09a}, BALQSOs have \mbox{$BI_0>0\,\mathrm{km\,s^{-1}}$}. We collected the values of $BI_0$ for our sample from \cite{Gibson09a}. Symbols in Fig.~\ref{fig1} are color coded according to their balnicity indices. Fig.~\ref{fig2} presents the X-ray weakness as a function of $BI_0$. 
Objects in our sample with strong BAL features ($BI_0>1000\,\mathrm{km\,s^{-1}}$) are indeed typically X-ray weak ($-$\daox=0.2--0.8). Sources with lower values of $BI_0$ show a large scatter of X-ray weakness.  \cite{Gibson09a} do not report uncertainties on $BI_0$, which are dominated by the continuum emission measurement. We estimated its uncertainty from \cite{FilizAk13}, who studied a subsample of \cite{Gibson09a}, and used the median value to approximate the uncertainties on $BI_0$ in  Fig.~\ref{fig2}.

\subsection{Expanding the sample to mini-BALQOS}\label{miniBALQSOs}

Our four targets happened to have quite weak BALs, as measured by the balnicity indices ($BI_0\approx100\,\mathrm{km\,s^{-1}}$) and are placed in the weak-BAL regime, where the X-ray weakness shows a large scatter. It is therefore worth extending the parameter space to related objects with even weaker features. We thus considered the sample of mini-BALQSOs of \cite{Wu10} that satisfy the quality cut described in \S~\ref{lit_sample}. These objects are defined to have $BI_0=0\,\mathrm{km\,s^{-1}}$ and absorption index $AI>0\,\mathrm{km\,s^{-1}}$, where 

\begin{equation}
AI=\int_0^{29000}(1-f_v)C'dv,
\end{equation}
and $C'= 1$ when the velocity width is at least $1000\,\mathrm{km\,s^{-1}}$ and the absorption trough falls at least 10\% below the continuum; $C'=0$ otherwise. Mini-BALQSOs thus represent an intermediate class of objects between BALQSOs and narrow absorption line QSOs (NALQSOs). The \cite{Wu10} mini-BALQSOs show normal levels of X-ray emission (i.e. \daox=0; Fig.~\ref{fig2}), restricting the scatter of \daox\, at low values of $BI_0$. They are also characterized by high accretion rates (log$\lambda_{Edd}=[-0.6,0.2]$), and indeed occupy the same region of the \daox-$\lambda_{Edd}$ plane as our two targets of rapidly accreting BALQSOs with low $BI_0$ (Fig.~\ref{fig1}), thus reinforcing our conclusion that the X-ray weakness is not apparently driven by the Eddington ratio. In fact, repeating the Spearman's  $\rho$ and Kendall's  $\tau$ tests with the addition of this sample of mini-BALQSOs returns even lower probabilities in favor of a correlation between $\lambda_{Edd}$ and \daox; i.e. 0.343 and 0.07, respectively. On the other hand, the anti-correlation between $\Delta\alpha_{ox}$ and $\mathrm{BI}_0$ is confirmed with high significance ($>0.999$ for both tests).

%This result is quite puzzling, since to first order the X-ray flux must be suppressed more strongly at higher accretion rates to avoid the overionization of the outflowing material. 

\subsection{Discussion}\label{Discussion}
According to Fig.~\ref{fig2} and the statistical tests run in the previous section, prominent BAL features are good proxies of X-ray weakness in BALQSOs, irrespective of their Eddington ratio. While the fraction of BALQSOs appears to depend on the Eddington ratio \citep[e.g.][]{Ganguly07}, the lack of a clear correlation between $\Delta\alpha_{ox}$ and $\lambda_{Edd}$ suggests that the observable X-ray-to-UV relative emission does not depend on the accretion rate. Since the optical/UV SEDs of quasars do not show evidence for a strong variation with Eddington ratio \citep[e.g.][]{Scott14}, this means that also the X-ray part of the SED does not vary strongly, at least for optically selected BALQSOs.
Our entire sample of 34 BALQSOs spans a relatively narrow range of luminosity ($46.3<\mathrm{log}L_{\mathrm{bol}}\lesssim47.8$; see Tab.~\ref{tab_sample}), and thus we do not expect strong evolution in luminosity to affect this result.

We note that sources with strong BAL features ($BI_0>1000\,\mathrm{km\,s^{-1}}$) show a hint of an anti-correlation between \daox\, and $\lambda_{Edd}$: the Spearman's $\rho$ test and the generalized Kendall's $\tau$ test return a probability of anti-correlation of $\approx0.94$ and $\approx0.96$, respectively, considering only the 14 BALQSOs with $BI_0>1000\,\mathrm{km\,s^{-1}}$, although, the small sample size and the the narrow $\lambda_{Edd}$ regime ($-1\lesssim \mathrm{log}\lambda_{Edd}\lesssim-0.4$) spanned prevents us from drawing solid conclusions. Most of the scatter in Fig.~\ref{fig1} is indeed due to BALQSOs with low $BI_0$. We speculate that the putative anti-correlation may thus be in place for BALQSOs with powerful and efficient outflows. To check this hypothesis, larger samples of BALQSOs with strong BAL features and accurate measurements of SMBH mass must be observed in X-rays. Weaker and less-massive winds may instead be present even if the X-ray emission is not significantly suppressed along the line of sight (see also, e.g. \citealt{Gibson09b, Hamann13}).  For instance, stochastic events, such as a local overdensity of the disk or the intervening of a dense gas cloud, may provide the needed screening against the ionizing radiation to locally allow the acceleration of a wind, which would thus be detected through absorption features much weaker than in the case of global, massive outflows, but would not cause significant X-ray absorption along the line of sight.

In order to check if our results are sensitive to the particular quality cuts we imposed in \S~\ref{lit_sample}, we repeated the analysis with a more conservative selection requiring $SNR_{\mathrm{Mg~II}}>10$ and $SNR_{\mathrm{C~IV}}>10$. This conservative sample consists of 19 BALQSOs, including our 4 newly observed sources. The results hold, although with lower significance, due to the smaller sample size.

Finally, we note that the observed scatter of any intrinsic relation can be increased by orientation effects. In fact, orientation is known to play a non-negligible role in the determination of physical properties of quasars \citep[e.g.][]{Jarvis06,Shen14}, and, in particular, of black-hole virial mass estimates \citep[e.g.][]{Runnoe13}. Similarly, observational properties of BALQSOs, such as the strength and velocity of the absorption features, may differ along different lines of sight \citep[e.g.][]{FilizAk14}.

\section{Summary}

We investigated the existence of a possible relation between the observed X-ray weakness (\daox) and the Eddington ratio $\lambda_{Edd}$ in a sample of 34 BALQSOs. Such a trend could help alleviate the overionization of outflowing material necessary to produce a disk wind, if line-driven radiation pressure is the main accelerating mechanism. Moreover, both theoretical and observational findings suggest a change of the level of X-ray production in quasars approaching the Eddington limit. However, we did not find evidence for a strong anti-correlation between \daox\, and $\lambda_{Edd}$.
Instead, the strength of the BAL features appears to be a better tracer of \daox\, than the Eddington ratio. Our results are confirmed also by considering a sample of mini-BALQSOs collected from the literature. 
Future X-ray observations of larger samples of BALQSOs with strong absorption features and accurate measurements of black-hole mass are needed to check if the anti-correlation between \daox\, and $\lambda_{Edd}$ is in place at least in the subpopulation with strong and massive winds, for which a stronger suppression of the X-ray emission may be required to accelerate the outflows efficiently.

\section*{Acknowledgments} \vspace{0.2cm}
We thank the anonymous referee for their comments and suggestions which improved the analysis and the presentation of the results.
 FV and WNB thank NASA XMM-Newton grant 80NSSC18K0487 and the NASA ADAP program for support. BL acknowledges financial support from the National Key R\&D 
 Program of China grant 2016YFA0400702 and National Natural Science 
 Foundation of China grant 11673010. We warmly thank J. Runnoe and Z. Shang for useful discussion on UV SED shape of quasars.
\\
\bibliography{biblio}

% \end{thebibliography}
% 
% 
% \bsp
% 
 
% 

%

\end{document}